\documentclass[a4paper,twoside]{article}
\usepackage{enumitem}
\usepackage{epsfig}
\usepackage{subcaption}
\usepackage{calc}
\usepackage{amssymb}
\usepackage{amstext}
\usepackage{amsmath}
\usepackage{amsthm}
\usepackage{multicol}
\usepackage{pslatex}
\usepackage{apalike}
\usepackage{subcaption}
\usepackage{hyperref}
\usepackage{xurl}
\usepackage{breakurl}
\usepackage{algorithm2e}
\usepackage[bottom]{footmisc}
\usepackage{SCITEPRESS}     % Please add other packages that you may need BEFORE the SCITEPRESS.sty package.

\begin{document}

\title{Analysing Multidisciplinary Approaches to Fight Large-Scale Digital Influence Operations}

\author{\authorname{David Arroyo\sup{1}\orcidAuthor{0000-0001-8894-9779}, Rafael Mata Milla\sup{1}\orcidAuthor{0009-0005-5009-1582}, Marc Almeida Ros\sup{1}\orcidAuthor{0009-0006-0827-1747}, Nikolaos Lykousas\sup{2}\orcidAuthor{0000-0001-8874-1230}, Ivan Homoliak\sup{3}\orcidAuthor{0000-0002-0790-0875}, Constantinos Patsakis\sup{4}\orcidAuthor{0000-0002-4460-9331}, and Fran Casino\sup{4,}\sup{5}\orcidAuthor{0000-0003-4296-2876}}
\affiliation{\sup{1}Spanish National Research Council (CSIC), Spain.}
\affiliation{\sup{2}Data Centric, Romania.}
\affiliation{\sup{3}Brno University of Technology,  Czech Republic}
\affiliation{\sup{4}Athena Research Centre, Greece.}
\affiliation{\sup{5}Department of Computer Engineering and Mathematics, Rovira i Virgili University, Tarragona, Spain.}
\email{franciscojose.casino@urv.cat}
}

\keywords{
Disinformation, Cyber warfare, Social networks, Crime as a service, Campaigns}

\abstract{Crime as a Service (CaaS) has evolved from isolated criminal incidents to a broad spectrum of illicit activities, including social media manipulation, foreign information manipulation and interference (FIMI), and the sale of disinformation toolkits. This article analyses how threat actors exploit specialised infrastructures ranging from proxy and VPN services to AI-driven generative models to orchestrate large-scale opinion manipulation. Moreover, it discusses how these malicious operations monetise the virality of social networks, weaponise dual-use technologies, and leverage user biases to amplify polarising narratives. In parallel, it examines key strategies for detecting, attributing, and mitigating such campaigns by highlighting the roles of blockchain-based content verification, advanced cryptographic proofs, and cross-disciplinary collaboration. Finally, the article highlights that countering disinformation demands an integrated framework that combines legal, technological, and societal efforts to address a rapidly adapting and borderless threat.}

\onecolumn \maketitle \normalsize \setcounter{footnote}{0} \vfill

\section{\uppercase{Introduction}}
\label{sec:introduction}

Crime as a Service (CaaS) is an emerging framework wherein malicious actors market and distribute specialised tools, infrastructure, and expertise to those seeking to engage in illicit activity. In its earliest forms, CaaS revolved around hacking services, malware kits, and data theft. However, as online ecosystems have evolved, so have the offerings under this service-based model. A key development has been the transition from purely criminal operations (e.g., ransomware or phishing) to more insidious activities such as disseminating disinformation and manipulating public opinion~\cite{smith2020influence,patsakis2024malware}.

Mirroring the industrialisation of cybercrime, CaaS markedly lowers the barrier to executing complex influence operations by offering ready-made toolkits that replace the need for significant technical or organisational expertise \cite{jones2019disinfo}. This toolbox may include botnet rentals, false identity account sets, disinformation playbooks, or even direct partnerships with professional trolls, all of which can be employed to influence public opinion, harass adversaries, and introduce chaos into digital conversations~\cite{brown2021weaponization}. In doing so, these teams systematically exploit the algorithms of social networks and repurpose them to magnify divisive or provocative content~\cite{lee2020stateactor}.

Threat actors in disinformation campaigns must rely on structures and technical resources to achieve their objectives, which in the end imply an effect on human behaviour. Especially relevant is the entire set of tools, services, and platforms that make it possible to separate the cause of a foreign information manipulation and interference (FIMI) strategy from its consequences. As noted in works such as~\cite{10.1145/3199674}, there is a gradual division of tasks and specialisation of cybercrime tools and services globally. The latter also applies to illicit activities in fabricating, instrumentation, and exploiting forged or decontextualised content, which has intensified in the last 10 years with social media and artificial intelligence~\cite{bentley2024cultural}. Actually, there has been a Fordist transformation~\cite{hirsch1991fordist} of the disinformation ecosystem due to the increasing commoditization of cybercrime and cognitive warfare.

This article contributes a structured analytical framework that consolidates disparate technological, economic, and operational dimensions of large-scale digital influence operations. While prior work has examined individual mechanisms, botnets, generative AI, Open Source Intelligence (OSINT) manipulation, or cyber-proxy infrastructures, we integrate these elements into a coherent taxonomy of enablers and actors that clarifies how disinformation ecosystems industrialise and scale. Our discussion highlights not only the urgency of adopting secure provenance mechanisms, automated detection pipelines, and multidisciplinary investigative cooperation, but also the societal consequences of inaction: persistent information pollution, diminished institutional credibility, and increased susceptibility to geopolitical manipulation. First, we characterise the critical components that enable these campaigns, including the proliferation of AI-driven techniques and dual-use technologies. Next, we map out the evolving landscape of threat actors and their economic models, illustrating the challenges of attributing cross-border interference. Finally, we review current strategies for countering these sophisticated attacks and discuss how emerging cryptographic and collaborative frameworks can help safeguard public discourse.

\section{Mapping CaaS onto International Social Networks}

The structure of CaaS can be mapped through a network of specialised actors who provide distinct components of an illicit campaign. On the one hand, technical experts offer custom software, exploit kits, and automated services (such as bot creation and management) for rent or purchase. On the other hand, infrastructure providers provide bulletproof hosting and VPN-like services that shield these operations from detection or shutdown. In addition, propaganda strategists shape narratives, craft messaging, and coordinate with ``troll farms'' that propagate talking points across multiple platforms and media~\cite{johnson2018actors}.

Revenue streams within CaaS operations come from multiple sources. Some groups operate on a subscription model, charging monthly fees for continuous access to botnets, other types of controlled accounts \cite{yamak2018multiple} (e.g., sock puppet or compromised accounts) or ``disinformation campaign managers". Others take a percentage of the client's gains, especially where the goal is monetised engagement or reputational damage (for instance, short-selling stocks after crafting negative news of specific companies)~\cite{thompson2021social}. In more politically oriented scenarios, state-affiliated or state-sponsored entities are willing to pay large sums for overseas interference, with payments often routed through obscure transactions or cryptocurrency. The strong anonymity in these financial exchanges compounds the challenge of tracking and dismantling these networks~\cite{garcia2020algorithmic}.

The commercial value of disinformation hinges on social-media virality, amplified by platform engagement mechanisms that artificially elevate visibility.. This viral spread is further facilitated by the ecosystem of likes, shares, retweets, and comments, all of which can be artificially inflated through paid engagements, thus reinforcing the demand for CaaS offerings~\cite{brown2021weaponization}. Recent research highlights how commercial disinformation vendors market region-specific solutions, targeting linguistic and cultural nuances in different countries~\cite{ross2022transnational}.

Simultaneously, the disinformation supply chain leverages everyday users as amplifiers without awareness. Content that sparks emotional reactions is more likely to be shared, providing operators with a direct path into the mainstream conversation~\cite{smith2020influence}. In polarised societies, these operations intensify echo chambers by reinforcing existing user biases, fragmenting audiences into closed loops of misinformation~\cite{jones2019disinfo}. The global consequences of these campaigns, ranging from political disruption to public health crises, underscore the need for international cooperation to detect, attribute, and counteract them.

\section{Structure and Elements of Disinformation Campaigns}
The main structures, technical means of disinformation campaigns, and technologies that enable a dual use in this context are summarised in the following sections. %Figure \ref{fig:tech} highlights the main keywords related to technological means used in disinformation campaigns.

\subsection{Main Technical Means}

\begin{description}[style=unboxed,leftmargin=0cm]

\item[VPNs and Residential Proxies.] A VPN (Virtual Private Network) allows one to browse the Internet while preventing the interception of the information exchanged. Although VPN services do not guarantee anonymity, they can serve as an additional layer of security that protects the threat actor's true identity, increasing the difficulty of attribution. Investigating VPN-sourced IP addresses imposes significant coordination overhead on law enforcement, particularly when providers enforce strict no-logging policies. As will be highlighted later, there is a dual circumstance here: protecting privacy can make it more difficult to attribute illicit activities in the cyber domain and, specifically, within the FIMI ecosystem. Moreover, if the network infrastructure used is internationally distributed, the investigator's work requires coordination across different legal frameworks that are not always compatible and hinder the necessary collaboration.

In the case of proxy servers, a user connects to this service to avoid directly accessing an Internet service or platform. %The proxy server acts as an intermediary between the client and the final server, thereby adding an extra layer between them. 
Among the different types of proxy servers, anonymous and residential proxies are particularly relevant. In anonymous proxies, the client's IP address is masked, so the accessed platform or service does not learn the original address. This allows a cybercriminal or threat actor in a campaign to conceal their attack vector, although the degree of protection will depend on the server type. In the case of residential proxies, the IP addresses belong to real residential IP addresses of existing users who `lend' their IP addresses to service providers, rendering blocking mechanisms useless, as their actual traffic is benign. Yet, the origin of user consent of many of these service providers is often shady~\cite{Mehanna_2024}. Residential proxies can also facilitate access to geographically restricted information. Likewise, a threat actor could use them to mislead investigators about the actual origin of the threat by positioning themselves in third-party countries that may also have an interest in the spread of disinformation. The use of residential proxies is part of the arsenal deployed, for instance, by APT29. This group was particularly active in Ukraine before the 2014 crisis and resumed active phishing campaigns in late 2018. In the FIMI context, the connection between APT29 and Russia's diplomatic actions is especially important~\cite{kunkle2021apts}. The abuse of legitimate services to carry out cyberattacks has been commonly observed in deploying botnets and Command and Control (C2) systems~\cite{al2023abuse}. The instrumental use of any such platform's services to bypass institutional, organisational, or corporate security filters and controls should be considered highly critical~\cite{cloudfare}. Thus, whitelisted service providers, especially for the cloud, are often used for hosting and bypassing firewalls and geographical restrictions.

\item[Phishing and SIM Swapping]
Phishing has become one of the most pervasive threats, facilitating numerous crimes. Disinformation is no exception to this rule as threat actors use social engineering to lure their victims into performing an action that otherwise would not do. As a result, they may disclose credentials, install malicious apps, or perform transactions, to name a few. Thus, phishing is among the top tools to trick crucial people into granting them access to their systems and social accounts. SIM swapping has traditionally been associated with finance-related scams, but it plays a new role in FIMI. In addition to enabling total or partial access to a device or service, which could be used for cyber espionage or to gain access to other elements of interest in a victim's account, this technique could also facilitate the acquisition of legitimate dissemination means, such as reputable social media user accounts with a large set of followers. This technique has other implications because it enables both the dissemination of disinformation and the discrediting of the impersonated victim. As with SIM swapping, phishing can be used to gain access to a legitimate account, enabling the dissemination of information to a larger audience or the acquisition of accurate information that can be manipulated for disinformation purposes. In the case of David Satter, his email account was illegally accessed, enabling his emails to be modified to publish false information about a supposed U.S.-financed operation to destabilise Russia~\cite{hulcoopTaintedLeaksDisinformation2017}. The SMS phishing ecosystem has also gained particular relevance in recent times. The entire set of hosting services for phishing and social engineering campaigns offers an idea of the level of sophistication of such activities~\cite{nahapetyan2024sms}, as well as of the challenges related to the supervision of service providers and platforms under the European Digital Services Act (DSA)~\cite{dsa}. Collecting information on hosting services, identifying patterns in digital certificate creation through certificate transparency logs, and tracking the development kits used for phishing campaigns are crucial in formulating containment strategies against FIMI.

\item[Generative Artificial Intelligence and Impersonation (Human Spoofing).] In an era where a lot of the content is AI-generated to the point of realising the "Dead Internet Theory", generative AI plays a crucial role in human influence~\cite{Walter25}. Large Language Models (LLMs) can generate plausible and convincing information about the target subject for disinformation. They also enable massive content generation, leading to information overload (infoxication). AI can refine disinformation by making the information more persuasive, improving its quality, or tailoring it to the target audience. Although most LLMs are censored or have safeguards to prevent such behaviour, they can be easily bypassed through various techniques such as chain-of-thought prompting and using uncensored or protected models~\cite{barmanDarkSideLanguage2024}. In case of requiring a source image for decontextualisation, one can now simply fabricate it with models like Flux, which, for example, do not contemplate censoring recognised individuals and where the difficulty of distinguishing generated images from real ones keeps growing. Through AI-based ``Human Spoofing'', disinformation actors can convincingly impersonate public figures for disseminating fabricated content or for deception and social engineering purposes. An example is the fake video call between the Mayor of Madrid and the supposed Mayor of Kyiv, which could be categorised as somewhere between a prank and hybrid warfare~\cite{twomeyDeepfakeVideosUndermine2023}.

\end{description}

\subsection{Dual-Use Technologies}
We define dual-use technologies as those that can be employed for benign use but may also be leveraged by threat actors for disinformation operations. Social media networks are a prime example; However, in this section, we focus on two primary infrastructures needed for conducting disinformation operations.

\begin{description}[style=unboxed,leftmargin=0cm]
    \item [Digital Marketing.] Digital marketing has become an essential tool for companies and organisations seeking to promote their products and services. Through strategies like Search Engine Optimization (SEO) and social media advertising, brands can reach specific audiences efficiently and effectively. However, the same techniques and platforms used for commercial objectives can also be employed for influence and disinformation operations. Threat actors can leverage segmentation tools and data analytics to identify and target vulnerable audiences, disseminating false or manipulated information to influence opinions, behaviours, or political decisions~\cite{marwick2017media}. For example, state or non-state actors have launched disinformation campaigns on social media to plant discord, polarise society, or discredit candidates during election processes. By creating deceptive content and using bots and fake profiles, they can amplify messages to appear more legitimate or popular than they actually are. There are also organised groups funded by states that offer services for social media manipulation campaigns. This is the case of so-called troll and/or bot farms~\cite{hughes2021macedonian}, or the use of decommissioned military infrastructure to support cybercrime and cyberwarfare~\cite{bunker}. Moreover, micro-segmentation techniques or recommender systems~\cite{deldjoo2024fairness} allow specific messages to be tailored to particular groups, thus increasing the effectiveness of manipulation~\cite{o2021microtargeted}. Likewise, threat actors can exploit character encodings, e.g., invisible Unicode characters, to manipulate search engines' results to present their content in user searches~\cite{BoucherPSAC23}. 
    
    \item [Wikipedia.] This cooperative encyclopedia is one of the most visited websites worldwide, serving as an accessible and free information source for millions of people. Studies leveraging Wikipedia for monitoring extremist networks, historical reconstruction, or linguistic pattern analysis illustrate its utility as a high-quality data source when combined with rigorous methodology \cite{mesgari2015sum}. In this regard, while wikipedia serves as a Anyone's ability to contribute to its editing is both its principal strength and its Achilles heel. From the perspective of disinformation and influence campaigns, this capacity is open to exploitation by threat actors through malicious editing and content bias, coordinated editing attacks, manipulation of sources and references, or the inclusion of sources and references that appear legitimate but are, in fact, part of a disinformation campaign. Likewise, Internet archives have been used extensively by digital-forensics teams to document coordinated harassment, attribute coordinated influence activities, and track the evolution of online criminal infrastructures. Highlighting these constructive applications reinforces the argument that defensive strategies should aim not at restricting these platforms, but at reinforcing their resilience against manipulation while preserving their crucial societal function. Although most deceptions on Wikipedia are quickly detected and have minimal impact, a small number persists for a long time and garner many views. As underlined in ~\cite{mcdowell2020takes}, human readers tend to view short articles as misleading content when, in fact, more extensive articles are more prone to deception. Furthermore, the ability to evade moderation for malicious content is limited without specialised automated tools. Despite efforts to counter them, the previously discussed factors still make Wikipedia an ideal venue for spreading misinformation and disinformation.%\todo{what about the case where there are additions to the article, so the article is not to blame but minor edits? This should be added}\todo[inline]{FC: Is misinformation and disinformation also to be stated somewhere? perhaps by adding what you suggest we could say sth about the different ways to elaborate text in that line}\todo[inline,color=blue!10]{CP:I believe that we should add a line about that. People are deliberately editing existing Wikipedia articles for disinformation and misinformation, it's not only the new articles}
    
    \item [Open Source Intelligence.] OSINT has been a significant source of information for many years, thanks to its easy accessibility, broad availability, and lower economic cost than other sources. The proliferation of technology and the mass adoption of the Internet have transformed the way information is gathered and analysed. OSINT has become a critical tool for extracting relevant data from a vast ocean of publicly available information. Its ease of acquisition has enabled it to be actively exploited by intelligence agencies, organisations, and individuals, improving decision-making across multiple sectors. Although OSINT is widely used and accepted, it is crucial to recognise the potential for disinformation and manipulation within this domain. A threat actor could cause infoxication by producing artificial content that is erroneous or subtly false, as well as by fabricating biased information that may be misinterpreted by adversaries or competitors~\cite{flamer2023enemy}. %One example is the excessive reliance that Hamas allegedly placed on OSINT, which Israel capitalised on by distributing messages in national media to shape a narrative intended to mislead Hamas' intelligence service~\cite{flamer2023enemy}. We should also consider the fight to dominate the prevailing narrative regarding a conflict by using OSINT as a legitimising source and, potentially, as a conduit for introducing manipulated information to shape public opinion.
    
    \item [Internet Archive.] Web archiving services like the Wayback Machine (from the Internet Archive) or Archive. It allows websites to be stored automatically or on demand, enabling users to access information even if it has been deleted or modified without visiting the original site. These services are crucial in preserving digital history, facilitating academic research, and providing access to information that could otherwise be lost. However, while their legitimate purpose is clear, malicious actors have been observed using this technology to propagate retracted or erroneous information or outright disinformation. Other actions include accessing media outlets opposed to their ideologies to reduce the advertising revenue of these outlets, evading censorship measures when disseminating \emph{disinformative} content on social networks, and capturing posts and news that might be deleted due to controversies~\cite{zannettouUnderstandingWebArchiving2018}.
\end{description}

\section{Discussion, Challenges, and Issues}
The state sponsorship of many disinformation campaigns hinders generic solutions, e.g., the establishment of international regulatory standards. The latter leaves a fragmented landscape that necessitates the cross-border intelligence sharing and adoption of domestic and regional technological measures. Thus, we summarise the insights from the previous analysis, highlighting the implications of disinformation campaigns and providing a perspective on future steps. The main discussion points focus on the technological advances needed to mitigate the threat of disinformation. A summary of the main strategies and the potential solutions they provide is shown in Table \ref{tab:summ}.

\begin{table}[t]
\centering
\scriptsize
\caption{Key strategies for countering disinformation and the challenges they address.}
\begin{tabular}{p{0.25\columnwidth} p{0.65\columnwidth}}
\hline
\multicolumn{1}{c}{\textbf{Strategies}} & \multicolumn{1}{c}{\textbf{Solved Challenges}} \\
\hline
  Identification of Actors, Tools, and Intermediaries in the FIMI Domain
&
  \vspace{-0.5\baselineskip}
\begin{itemize}[leftmargin=*,nosep]
  \item Clarifies and unifies investigative approaches
  \item Facilitates cross-border cooperation
  \item Reduces fragmentation in threat analysis
\end{itemize}
\vspace{-\baselineskip} % trims bottom space
\\
\hline
  Digital Signatures and Secure Protocols for Content Verification
&
\vspace{-0.5\baselineskip}
\begin{itemize}[leftmargin=*,nosep]
  \item Confirms the authenticity of digital content
  \item Mitigates the spread of manipulated media
  \item Builds trust in digital exchanges
\end{itemize}
\vspace{-\baselineskip}
\\
\hline
  Preserving Integrity During Content Modifications
&
\vspace{-0.5\baselineskip}
\begin{itemize}[leftmargin=*,nosep]
  \item Maintains verifiable links after cropping or editing
  \item Ensures chain-of-custody for processed media
  \item Prevents untraceable transformations
\end{itemize}
\vspace{-\baselineskip}
\\
\hline
  Automated Disinformation Detection Using AI and Big Data
&
\vspace{-0.5\baselineskip}
\begin{itemize}[leftmargin=*,nosep]
  \item Deals with massive scales of misinformation
  \item Provides near real-time classification
  \item Minimises dependency on manual reviewers
\end{itemize}
\vspace{-\baselineskip}
\\
\hline
  Experts in News Verification
&
\vspace{-0.5\baselineskip}
\begin{itemize}[leftmargin=*,nosep]
  \item Adds human oversight for context-rich analysis
  \item Counteracts deepfakes and manipulation
  \item Bridges purely algorithmic gaps
\end{itemize}
\vspace{-\baselineskip}
\\
\hline
    Blockchain and Immutable Ledgers
&
\vspace{-0.5\baselineskip}
\begin{itemize}[leftmargin=*,nosep]
  \item Creates tamper-evident records for media
  \item Strengthens provenance tracking
  \item Reduces single points of failure in verification
\end{itemize}
\vspace{-\baselineskip}
\\
\hline
  Multidisciplinary Collaboration in the Fight Against Disinformation
&
\vspace{-0.5\baselineskip}
\begin{itemize}[leftmargin=*,nosep]
  \item Enhances coordination across law enforcement
  \item Integrates policy, technology, and social sciences
  \item Fosters coherent international frameworks
\end{itemize}
\vspace{-\baselineskip}
\\
\hline
  Enhancing Cyber Attribution
&
\vspace{-0.5\baselineskip}
\begin{itemize}[leftmargin=*,nosep]
  \item Identifies threat actors behind manipulations
  \item Helps counter false-flag operations
  \item Improves global cooperation for enforcement
\end{itemize}
\vspace{-\baselineskip}
\\
\hline
\end{tabular}
\label{tab:summ}
\end{table} 

Given the crucial role of fake accounts in the dissemination of disinformation, their prompt detection is vital. However, this can be done with the correlation of various heterogeneous information. For instance, user interaction, post timing, social network connections, shared content, etc., can be used to determine whether the user is a human or a bot. 

% \todo[inline]{Botnet detection? Detection of fake/bot accounts}
%Development of Standards and Procedures for the Identification of Actors, Tools, and Intermediaries in the FIMI Domain. 
From the standpoint of attributing campaigns and foreign interference actions, the association between states and actors is both a technological and methodological challenge~\cite{eeas}. The sophistication and specialisation in the products, services, and platforms for creating and distributing fabricated content further intensify the multi-faceted nature of the concept of a cyber-proxy~\cite{borghard2016can}. Risk and threat analysis concerning different types of proxies in the cyber sphere requires the development of procedures for the identification, annotation, and sharing of evidence and intelligence at a transnational level. Thus, at the European level, it would be worth integrating technological solutions derived from theoretical frameworks for risk analysis and FIMI action attribution into the Information Sharing and Analysis Centers (ISACs) network.

%Digital Signatures and Secure Protocols for Content Verification. 
Disinformation campaigns can be effectively countered by adopting secure protocols and digital signatures that authenticate the content's origin and integrity. Techniques like Trusted Execution Environments (TEE) and Zero-Knowledge Proofs (ZK-SNARKs) are emerging as viable options for verifying the authenticity of information. These cryptographic techniques ensure that the original authorship of digital content remains intact, thus reducing the likelihood of uncontrolled dissemination of manipulated or fabricated material. Implementing these protocols in environments where Generative AI can fabricate disinformation will be crucial in the future.

A well-known case study in the fight against disinformation using zk-SNARKs is the protection of digital images. Cameras require a secure element to sign captured images or videos. This secure element is essentially a tamper-proof signing mechanism, often considered similar to a TEE. Several companies have contributed to a standard called C2PA (Coalition for Content Provenance and Authenticity)~\cite{c2pa_spec}, which does not necessarily mandate secure hardware but instead focuses on binding multimedia content to reliable metadata (e.g., geolocation). C2PA allows users to balance privacy and authenticity by, for instance, attaching geolocation to an image while selectively revealing the information they choose to share. For example, a 30 MP photo and its corresponding signature, can prove that an authenticated camera captured the photo in a particular place. This mechanism helps to combat disinformation, such as the spread of fake photos in conflict zones. Possible attacks, such as photographing a printed image, remain a separate vector that must be addressed independently.

%Preserving Integrity During Content Modifications. 
News agencies and TV networks often modify the original photographs and videos before publishing them. Operations like cropping, resizing, or converting to grayscale result in losing the original linkage between the signed medium and its metadata. This creates a significant challenge for maintaining content integrity throughout the modification process. To address this, zk-SNARKs can be employed to preserve the linkage between the original image and its modifications. By defining a circuit that represents the transformations applied to the original medium, it is possible to keep the signature and linkage intact, even after altering the content. Various authors have discussed this approach~\cite{cryptoeprint:2024/1066}, although one of the main obstacles was the substantial computational overhead required to generate zk-SNARK proofs. Their method demanded large amounts of memory, up to 64GB, just to generate a single proof, hindering its broad adoption. More recent research~\cite{della2024trust} optimised this approach by using a divide-and-conquer principle, %, proposing to split an image into tiles and modify the C2PA protocol to sign these aggregated tiles using a Merkle tree structure. This method allows the generalisation of zk-SNARK proofs for individual tiles, substantially reducing the computational burden. As a result, this optimised approach 
enabling the generation of zk-SNARK proofs even on mainstream hardware (e.g., with 4GB of RAM).

%While methods for static images have shown promising results, video content poses a significantly greater computational challenge, making zk-SNARK proof generation much slower and more resource-intensive. Research is underway to parallelise zk-SNARK proof generation using GPU hardware, offering up to a 4x speed improvement. However, these efforts still do not meet the computational efficiency necessary for real-time or large-scale video processing.

%Automated Disinformation Detection Using AI and Big Data. 
LLMs can generate large volumes of plausible false content, intensifying existing information-overload problems and complicating detection efforts~\cite{xu2023combating}. Addressing this overload will require both technological solutions (e.g., improved filtering algorithms and automated verification systems) and educational efforts to enhance media literacy across society. In this context, automating disinformation detection via AI-based systems that leverage semantic correlations and Natural Language Processing becomes crucial. Still, challenges remain in terms of scalability and precision. Moreover, maintaining continuously updated databases to track known sources of disinformation will play a critical role in the effectiveness of these AI systems~\cite{mansurova2024qa}.

%Experts in News Verification. 
While automated systems play an important role, expert involvement in verifying content accuracy and legitimacy cannot be overlooked. The role of fact-checkers, journalists, and domain experts must be strengthened to counteract the increasing influence of AI-generated disinformation. Incorporating expert feedback into AI systems to detect and flag false content could result in a robust hybrid system~\cite{mahmud2023toward}. Moreover, the term ``expert'' itself requires definitions that can be used to identify the proper personnel along with their skill sets and behaviours in diverse informational contexts, constituting a research line of its own.

%Emerging Technologies in Protecting Verified Content: Blockchain and Immutable Ledgers. 
Proper curation of news by experts, combined with efficient annotation and distribution, requires procedures guaranteeing custody and integrity. Blockchain technology offers another potential solution to counter disinformation by creating immutable digital records of the contents published. This would allow information to be tracked as it spreads across platforms, ensuring that any alterations or manipulations of the original content are visible, traceable, and verifiable. Such an approach could be especially effective on news and social media platforms, where false information can spread quickly~\cite{fraga2020fake}.

%Multidisciplinary Collaboration in the Fight Against Disinformation. 
A threat-modeling perspective in cybersecurity can help better characterise the profiles of attackers in disinformation campaigns, including their attack patterns, preferred targets, and the techniques they use~\cite{mirza2023tactics}. However, this requires a collaborative approach that integrates multiple disciplines. Cybersecurity experts, law enforcement agencies, data analysts, psychologists, and semantic analysts must work together to develop effective countermeasures. This interdisciplinary collaboration should also extend to legislators, ensuring that the legal framework evolves alongside technological advancements~\cite{casino2025unveiling}. Protocols designed to detect and mitigate disinformation should be grounded in legal standards, providing both privacy and security~\cite{casino2022sok}.

%Challenges in Cyber Attribution. 
One of the persistent limitations in fighting disinformation is cyber attribution. Identifying the source of disinformation, particularly in cases involving state-sponsored campaigns, requires sophisticated tools and international cooperation. While advancements have been made, especially through the use of AI and OSINT, the evolving tactics of malicious actors make attribution increasingly challenging. False-flag operations, anonymising internet infrastructures (e.g., VPNs, proxies), and using encrypted communication channels hinder the ability to identify actors behind disinformation campaigns. Future efforts must focus on enhancing cyber attribution frameworks to hold perpetrators accountable at the international level~\cite{maesschalck2024gentlemen}.

\section{Conclusion}
This article explores the interplay between technology, cybercrime services, and disinformation campaigns, highlighting how malicious actors methodically exploit social networks and advanced AI tools to lessen trust on a global scale. By identifying similarities between disinformation campaigns and more traditional cyberattacks, we illustrated how attackers employ ready-made infrastructures such as botnets, AI-driven content generation, and social engineering, as integral parts of an increasingly commercialised threat. In turn, these services lower the barrier to entry for a range of adversaries, from cybercriminals pursuing financial gain to state-sponsored groups seeking sociopolitical disruption.

Effective defence thus requires multi-layered responses that unite technical innovation, legal frameworks, and cross-disciplinary collaboration, underscoring the need for deeper cooperation across law enforcement, intelligence agencies, and international governing bodies. Future work will explore the use of AI to leverage attribution and identification of disinformation campaigns through expert agents, merging human and machine capabilities to foster advancement in the field.

\section*{\uppercase{Acknowledgements}}

This work was supported by the European Commission as part of the project SAFEHORIZON (G.A. no. 101168562). This work was also supported by Brno University of Technology (FIT-S-23-8151), NextGenerationEU (Slovakia project 09I0503-V02-00057), and the Chips JU DistriMuSe project (G.A. no. 101139769). This work was supported by Ministerio de Ciencia, Innovación y Universidades, Gobierno de España (Agencia Estatal de Investigación, Fondo Europeo de Desarrollo Regional -FEDER-, European Union) under the research grant PID2024-158490OB-C31 ECEAAS, and by
Comunidad Autonoma de Madrid, CIRMA-CM Project (TEC-2024/COM-404). F. Casino was supported by the Spanish Ministry of Science and Innovation under the ``Ramón y Cajal” programme (RYC2023-044857-I).
The content of this article does not reflect the official opinion of the European Union. Responsibility for the information and views expressed therein lies entirely with the authors.

\bibliographystyle{apalike}
{\small
\bibliography{database}}

\end{document}